\documentclass[lettersize,journal]{IEEEtran}
\usepackage{amsmath,amsfonts}
\usepackage{algpseudocode}
\usepackage{algorithm}
\usepackage{array}
\usepackage[caption=false,font=normalsize,labelfont=sf,textfont=sf]{subfig}
\usepackage{textcomp}
\usepackage{stfloats}
\usepackage{url}
\usepackage{verbatim}
\usepackage{graphicx}
\usepackage{cite}
\usepackage{adjustbox}
\usepackage{float}
\usepackage{enumerate}
\usepackage{multicol, multirow, tabularx}
\usepackage[table]{xcolor}
\usepackage{amsthm}
\usepackage[para]{threeparttable}
\usepackage{subcaption}
\usepackage{booktabs}
\begin{document}

\title{Cryptanalysis via Machine Learning Based Information Theoretic Metrics} 

\author{ 
        Benjamin D. Kim, Vipindev Adat Vasudevan, Rafael G. L. D'Oliveira, \\ Alejandro Cohen, Thomas Stahlbuhk, and Muriel Médard% <-this % stops a space
%\thanks{This paper was produced by the IEEE Publication Technology Group. They are in Piscataway, NJ.}% <-this % stops a space
%\thanks{Manuscript received April 19, 2021; revised August 16, 2021.}
\thanks{Benjamin D. Kim is with the Department of Electrical and Computer Engineering, University of Illinois Urbana-Champaign, Champaign, IL 61820 USA (e-mail: bdkim4@illinois.edu). Vipindev Adat Vasudevan, and Muriel Médard are with the Department of Electrical Engineering and Computer Science, Massachusetts Institute of Technology, Cambridge, MA 02139 USA (e-mail: vipindev@mit.edu; medard@mit.edu). Rafael G. L. D'Oliveira is with the School of Mathematical and Statistical Sciences, Clemson University, Clemson, SC 29631 USA (e-mail: rdolive@clemson.edu). Alejandro Cohen is with the Technion Faculty of Electrical Engineering, Technion Israel Institute of Technology, Haifa 32000, Israel (e-mail: alecohen@technion.ac.il). Thomas Stahlbuhk is with MIT Lincoln Laboratory, Lexington, MA 02421 USA (e-mail: thomas.stahlbuhk@ll.mit.edu).}
\thanks{Part of this work was previously published in ICASSP 2024 \cite{kim2024crypto}.} %but is extended to a much more comprehensive analysis and includes more results.}
}



\maketitle

\begin{abstract}

    % We explore integrating the field of machine learning (ML) with the field of cryptanalysis. Interestingly, the fields of ML and cryptanalysis share a common goal of creating a function, based on a given set of inputs and outputs. However, the approaches and methods in doing so vary vastly between the two fields. In this paper, we propose two novel (applications of?) machine learning algorithms that can be applied in a known plaintext setting to perform cryptanalysis on any cryptosystem. In particular, these algorithms utilize information theoretic metrics to perform ML-based distribution estimation. We use mutual information neural estimation to calculate a cryptosystem’s mutual information leakage, and a BCE classification to model an indistinguishability under chosen plaintext attack. These algorithms can be readily applied in an audit setting to evaluate the robustness of a cryptosystem and the results can provide a useful empirical bound. We evaluate the efficacy of our methodologies by empirically analyzing several encryption schemes. Furthermore, we extend the analysis to novel network coding-based cryptosystems and provide other use cases for our algorithms.

    The fields of machine learning (ML) and cryptanalysis share an interestingly common objective of creating a function, based on a given set of inputs and outputs. However, the approaches and methods in doing so vary vastly between the two fields. In this paper, we explore integrating the knowledge from the ML domain to provide empirical evaluations of cryptosystems. Particularly, we utilize information theoretic metrics to perform ML-based distribution estimation. We propose two novel applications of ML algorithms that can be applied in a known plaintext setting to perform cryptanalysis on any cryptosystem. We use mutual information neural estimation to calculate a cryptosystem’s mutual information leakage, and a binary cross entropy classification to model an indistinguishability under chosen plaintext attack (CPA). These algorithms can be readily applied in an audit setting to evaluate the robustness of a cryptosystem and the results can provide a useful empirical bound. We evaluate the efficacy of our methodologies by empirically analyzing several encryption schemes. Furthermore, we extend the analysis to novel network coding-based cryptosystems and provide other use cases for our algorithms. We show that our classification model correctly identifies the encryption schemes that are not IND-CPA secure, such as DES, RSA, and AES ECB, with high accuracy. It also identifies the faults in CPA-secure cryptosystems with faulty parameters, such a reduced counter version of AES-CTR. We also conclude that with our algorithms, in most cases a smaller-sized neural network using less computing power can identify vulnerabilities in cryptosystems, providing a quick check of the sanity of the cryptosystem and help to decide whether to spend more resources to deploy larger networks that are able to break the cryptosystem. %Furthermore, we extend the analysis to novel network coding-based cryptosystems and provide other use cases for our algorithms.

    %evaluate a cryptosystem to address the rising concerns about the capability of adversaries to break computationally hard cryptosystems in a post-quantum era
    %to do: Appendix, Additional Citations throughout work.

\end{abstract}
\begin{IEEEkeywords}
Cryptography, Mutual Information, Entropy, Classification, Plaintext Attack
\end{IEEEkeywords}
\section{introduction}
\label{introduction}
In the modern era, the importance of digital communication and storage has increased exponentially, making security a paramount concern for individuals, businesses, and governments alike. As our reliance on digital platforms grows, ensuring reliable communication over inherently unreliable channels has become a critical challenge. To address this, cryptosystems have long been employed as the cornerstone of secure digital communication, providing methods to encrypt and protect sensitive information from unauthorized access. However, the landscape of cybersecurity is constantly changing. With the rapid advancement of hardware technologies and ever-increasing computational capabilities, cryptosystems that were considered to be secure are now at a high risk of being compromised \cite{mavroeidis2018impact}. This underscores the need for continuous innovation in encryption methods and security protocols as well as their verification. As quantum computing looms on the horizon, promising to render many current encryption techniques obsolete, there is an urgency to develop new, post-quantum cryptographic solutions \cite{chen2016report,bos2018crystals,prest2020falcon}. Furthermore, the advancements in artificial intelligence and machine learning algorithms have found their use in securing systems as well as breaking them \cite{dubrova2023breaking, wenger2022salsa}. With the heavy computational power and strong learning algorithms, cryptosystems can be learned over time and attackers can predict the encrypted data with high accuracy. The future of secure digital communication hinges on our ability to stay ahead of these technological advancements and create cryptosystems that can withstand the computational power of tomorrow.

Throughout their history, cryptographic protocols and systems have been designed to stay secure against attackers with some limitations on their computational strength \cite{stinson2018cryptography}. However, as computational power increases over time, these protocols are likely to be broken. AI/ML have accelerated this process due to their state-of-the-art pattern recognition capabilities, and can act as a potentially powerful aid to attackers. The importance of constant or regular monitoring or auditing of a security protocol has increased significantly with these developments in computational power \cite{gohr2019improving, wenger2022salsa, dubrova2023breaking}. %Interestingly, the fields of machine learning and cryptanalysis share a common goal of creating a function, based on a given set of inputs and outputs. However, the approaches and methods in doing so vary vastly between the two fields. 

%The importance of constant or regular monitoring or auditing of a security protocol has increased significantly with these developments in computiation power. 

Security auditing has become broadly studied in adjacent fields to cryptography, such as differential privacy \cite{dwork2006differential, dwork2008differential}. Rather than providing a theoretical verification of security, an empirical lower bound is estimated. This valuable empirical approach allows for practical evaluations of security in real-world applications, where formal theoretical guarantees may be difficult to apply. This has led to differential privacy being implemented in several practical settings, by companies such as Apple \cite{tang2017privacy}, Uber \cite{near2018differential}, and the U.S. Census Bureau \cite{abowd2018us}, etc. The application of ML algorithms has also been found useful in this auditing process \cite{nasr2023tight}. However, auditing on large datasets has been proven to be computationally expensive, as it requires running estimation algorithms hundreds of times. More recent advances in differential privacy have proposed auditing with fewer runs \cite{nasr2023tight,steinke2024privacy}. The empirical bounds efficiently obtained in these algorithms are crucial to ensure proper privacy.

% Many of these approaches focus on evaluating the security of cryptosystems based on information theoretic metrics such as mutual information leakage. 

% \textcolor{yellow}{probably expand this paragraph to give more details about auditing SOTA}

A cryptosystem can be said to achieve perfect secrecy if the mutual information (MI) between the ciphertext and the plaintext is zero. The assumption of perfect secrecy implies that one is unable to gain any information from the ciphertext about the plaintext. However, achieving perfect secrecy in a practical setting is extremely challenging \cite{shimeall2013introduction}. Most of the computationally secure cryptosystems relax some of the assumptions in the perfect secrecy condition to achieve practical security. Furthermore, modern cryptography depends on the hardness of solving a computational problem without a particular element, the key. The systems acknowledge that the systems are only secure up to a point where you can not do a particular number of operations within the required time and generally quantify this as `b-bit secure. However, with the evolution of quantum computing and larger computational capabilities, along with the learning algorithms, this notion of security may not be enough to ensure that the cryptosystems have a lifespan that is desired. It is essential to define the level of security based on the information-theoretic metrics that provide guaranteed performance against any adversary. However, it is extremely challenging to calculate such metrics, such as MI, for complex cryptographic protocols that use different permutation and combination operations to achieve the desired level of computational security.

Advancements in the machine learning domain have also led to some interesting developments in this direction as well. Even though evaluating MI between two multi-dimensional random variables remains a difficult problem, estimating a lower bound for the same using stochastic gradient descent (SGD) over their samples seems to be practical \cite{belghazi2018mine}. The mutual information estimation using neural networks has found its application in evaluating the MI leakage between the plaintext and ciphertext \cite{kim2024crypto} as well as many other privacy and information leakage evaluations \cite{zhang2022privacy,razeghi2023bottlenecks,rodriguez2021variational,esfahanizadeh2023infoshape,kale2024texshape}. Moreover, distribution estimation for classifying tasks has also been significantly enhanced by machine learning techniques. Neural networks can be used to estimate a distribution, optimizing with SGD, and capturing complex dependencies between variables. These developments enable more precise and efficient modeling, which have had several applications in the information security domain as well, and specifically can have powerful implications in the field of cryptography. If an adversary is able to correctly distinguish between ciphers by observing different plaintext classes, several cryptography metrics would deem the cryptosystem no longer secure. Both these methodologies can also be extended to a continuous monitoring system for evaluating the security level of a cryptosystem. 

\subsection{Main Contributions} In this work, we propose a new hybrid machine learning-based approach to audit the security level of cryptosystems and provide an empirical bound through MI estimation and chosen plaintext attack classification. Our results show that the mutual information estimation using neural networks identifies the information leakage efficiently and the binary classification model accurately differentiates between the schemes that are indistinguishability under chosen plaintext attack (IND-CPA) secure and the schemes that are not IND-CPA secure (defined in \ref{INDCPA_defn}). Furthermore, our small-sized neural networks can predict the vulnerability of the cryptosystem using limited resources before deploying larger resources to crack the system. The primary contributions of our study are summarized as follows: 
\begin{itemize}
    \item We investigate the fundamental application of an ML-based aid in cryptanalysis, in a classical known plaintext setting. We extend and discuss the similarities between the two fields along with the potential implications.
    \item We introduce a novel algorithm for ML models, classifying ciphers and acting as an adversary in an IND-CPA setting, utilizing entropy-based distribution estimation. We also extend the mutual information neural estimation analysis between ciphertexts and plaintexts seen in \cite{kim2024crypto}. 
    \item We empirically demonstrate the effectiveness of our proposed algorithms and experiment on several widely-used cryptosystems, also performing a comprehensive analysis of our results. 
    \item We demonstrate the relevance of information-theoretic metrics in our cryptanalysis framework and their direct impact on improving the effectiveness of the employed cryptanalytic techniques.
\end{itemize}

The rest of the paper is organized as follows. Section \ref{sec:definitions} presents relevant background and definitions both in the cryptography and ML domain. Section \ref{methodology} presents our main methodologies and algorithms used. Section \ref{results} contains our experimental results and Section \ref{discussion} contains a discussion on our work.

%Structure of the paper.

% Within the past decade Artificial Intelligence and Machine Learning (ML) have become profoundly influential
% The fields of machine learning and cryptanalysis share a common goal of creating a function, based on a given set of inputs and outputs. However, the approaches and methods in doing so vary vastly between the two fields. In this work we present and analyze new methods to gauge cryptosystems security with information theoretic measure through machine learning.

\section{Preliminaries and Definitions}
\label{sec:definitions}

This section provides an overview of some of the underlying concepts from the fields of cryptology and information theory that our proposed approaches are built on. It elaborates on the definitions and standard approaches in the evaluation of secure encryption schemes, and then introduces relevant metrics from information theory we use later on to perform a cryptanalysis.

\subsection{\textbf{Security Preliminaries}}
We first define the security preliminaries in this subsection starting with the concept of a cryptosystem. We then introduce the notion of chosen plaintext attacks, a standard for determining whether an encryption scheme is secure.

\subsubsection{\textbf{Cryptosystem Definition}}\label{cryptosystem_defn}
We begin by defining fundamental elements for encryption schemes or cryptosystems that are relevant to our analysis. In Section \ref{results}, we analyze our methods on both symmetric-key and public-key (asymmetric) schemes. A cryptosystem can be described as either symmetric or asymmetric, characterized by the following with a security parameter $\kappa$:

An asymmetric or public-key cryptosystem contains the following three algorithms:
\begin{itemize}
    \item A key generation algorithm $\text{Gen}(\kappa)$ which takes as input a security parameter $\kappa$ and generates a public key $p_k$ and a secret key $s_k$.
    \item An encryption algorithm $\text{Enc}(m, p_k)$ which takes as input a message $m$ belonging to some set of messages $\mathcal{M}$ and the public key $p_k$ and then outputs a ciphertext $c$ belonging to some set of ciphertexts $\mathcal{C}$.
    \item A polynomial-time decryption algorithm $\text{Dec}(c, s_k)$ which takes as input a ciphertext $c = \text{Enc}(m, p_k)$ and the secret key $s_k$ and outputs the original message $m$.
\end{itemize}

A symmetric cryptosystem contains the following three algorithms:
\begin{itemize}
    \item A key generation algorithm $\text{Gen}(\kappa)$ which takes as input a security parameter $\kappa$ and generates a secret key $k$.
    \item An encryption algorithm $\text{Enc}(m, k)$ which takes as input a message $m$ belonging to some set of messages $\mathcal{M}$ and the secret key $k$, and then outputs a ciphertext $c$ belonging to some set of ciphertexts $\mathcal{C}$.
    \item A polynomial-time decryption algorithm $\text{Dec}(c, k)$ which takes as input a ciphertext $c = \text{Enc}(m, k)$ and the same secret key $k$ and outputs the original message $m$.
\end{itemize}

% \begin{enumerate}
% \item \textbf{Key Generation:}
% \begin{itemize}
% \item \textit{Symmetric Cryptosystem:} The key generation algorithm $\mathrm{Gen}(\kappa)$ takes a security parameter $\kappa$ as an input and generates a single key $k$, which serves as the secret key.
% \item \textit{Asymmetric Cryptosystem:} The key generation algorithm $\mathrm{Gen}(\kappa)$ takes as input a security parameter $\kappa$ and generates a pair of keys: a public key $p_k$ and a secret key $s_k$.
% \end{itemize}

% \item \textbf{Encryption Algorithm:} The encryption algorithm $\mathrm{Enc}(m, k)$ takes as input a message/plaintext $m$ from a set of possible messages $\mathcal{M}$ and a key $k$ (which is either $p_k$ in the case of asymmetric cryptosystems or the single key $k$ in symmetric cryptosystems). It outputs a ciphertext $c$ belonging to a set of possible ciphertexts $\mathcal{C}$.

% \item \textbf{Decryption Algorithm:} The decryption algorithm $\mathrm{Dec}(c, k)$ is a polynomial-time algorithm that takes as input a ciphertext $c$ (which is $\mathrm{Enc}(m, k)$) and a key $k$ (which is either $s_k$ in the case of asymmetric cryptosystems or the single key in symmetric cryptosystems). It outputs the original message $m$.
% \end{enumerate}

With every computational cryptosystem comes the question of how secure it is, if an adversary was trying to gain information about the message $m$ from the cipher $c$. There are several ways to attack and exploit encryption schemes, including key based attacks \cite{bellare2003theoretical}, side-channel analysis \cite{standaert2010introduction} \cite{dubrova2023breaking}, energy consumption \cite{randolph2020power}, etc. In this work, we will focus on one of the most fundamental notions in the cryptographic field, IND-CPA. Stronger notions of indistinguishability attacks such as Chosen Ciphertext attacks (IND-CCA) have become standard for more recently proposed cryptosystems, but our classification framework presented in Section \ref{methodology} focuses on applications to IND-CPA.

\subsubsection{\textbf{Indistinguishability under Chosen-Plaintext Attack (IND-CPA)}}\label{INDCPA_defn}

A cryptosystem is IND-CPA secure if no Probabilistic Polynomial Time (PPT) adversary can distinguish between the encryption's of two chosen plaintexts with a non-negligible advantage. The IND-CPA security model is defined through the following game between an adversary \(\mathcal{A}\) and a challenger \(\mathcal{C}\):

\begin{itemize}
    \item The challenger generates the respective keys the encryption mechanism requires with $\mathrm{Gen}(\kappa)$ and if there is a public key $(\mathit{pk})$, it is sent to the adversary \(\mathcal{A}\).
    
    \item  The adversary \(\mathcal{A}\) is allowed to make polynomially many encryption queries. For each query, \(\mathcal{A}\) submits a plaintext \(m\) to the challenger, who responds with the ciphertext.
    
    \item  The adversary \(\mathcal{A}\) submits two distinct plaintexts \(m_0\) and \(m_1\) to the challenger \(\mathcal{C}\). The challenger randomly selects a bit \(b \in \{0,1\}\), and sends the ciphertext \(c^* = \mathrm{Enc}(\mathit{pk}, m_b)\) to \(\mathcal{A}\).
    
    \item  The adversary \(\mathcal{A}\) outputs a guess \(b'\) for the value of \(b\). The experiment outputs 1 if the guess is correct.
\end{itemize}

A cryptosystem is considered secure under the IND-CPA model if the adversary's probability of correctly guessing \(b\) is only negligibly better than guessing such that:

$$Pr [\text{CPA}_{\text{Game}}^{\mathcal{A}} = 1] \leq 1/2 + f(\kappa).$$

A function f: $\mathbb{N}  \rightarrow \mathbb{R}$  is considered negligible if for every positive integer $s$ there exists an integer $\kappa_s$ such that for all $\kappa > \kappa_s$, $f(\kappa) < 1/\kappa^s$ holds. Simply put, the adversary doesn't have a better chance than a random guess for which message the cipher came from.

%\textcolor{blue}{introduction for information theoretic metrics, connection between entropy and security, MI and security etc.)}
\subsection{\textbf{Information Theoretic Metrics}}
We now transition to the metrics and objectives used to perform our cryptanalysis in this work. Computationally secure cryptosystems base their security claims on the limitation of the adversary to perform complex mathematical operations without the key. However, information-theoretic approaches do not make that assumption of limited computational power \cite{liang2009information}; instead, they verify that the adversary can not differentiate its input from a random variable of the same length. In another way, the adversary would not be able to identify the plaintext due to a lack of useful information. There are different approaches and potential evaluation metrics to verify the decrypting capability of an adversary. We are primarily interested in two such parameters, Mutual Information and Binary Cross Entropy.

\subsubsection{\textbf{Mutual Information}}

Mutual information (MI) is an entropy-based measure, that quantifies the dependence and relationship between two variables. MI has an extensive history in the field of cryptography, allowing one to quantify information leakage in secure communication systems \cite{shannon1949communication}. The mutual information between random variables $X$ and $Y$ can be quantified as the following, where $H$ is the Shannon entropy
\[
I(X;Y) = H(X) - H(X|Y). 
\]

Mutual information can also be represented as the KL-divergence between the joint probability distribution between $X$ and $Y$, and the product of the marginals between $X$ and $Y$ 
\[
I(X;Y) = D_{KL}(P(X,Y) || P(X)P(Y)).  
\]

In this work, we take the Donsker Varadhan (DV) representation of the KL-divergence, where $\Omega$ is the product sample space of the distributions $P_1$ and $P_2$, and the supremum is taken over all functions $F$, that have a finite expectation. $F$ can be modeled as a neural network $F_{\phi}$, seen in \cite{belghazi2018mine}. $\phi$ can be computed and optimized through stochastic gradient descent \cite{goodfellow2016deep}, allowing one to estimate a lower bound of mutual information\footnote{All calculations in this work use the natural logarithm}
\begin{equation*}
D_{KL}(P_1 || P_2) = {\sup_{F: \Omega \to \mathbb{R}}} \mathbb{E}_{P_1}[F] - \log( \mathbb{E}_{P_2}[e^F] ).  
\end{equation*} 

When estimating MI with the above loss function in models, there have been issues in variance. This is remedied in \cite{choi2020regularized}, with an added stabilization term in the objective, leaving us with our final optimization objective to calculate a lower bound of MI 
\begin{multline}
    \label{MINE_objective}
        I_{\phi}(X;Y) = \mathbb{E}_{P(X,Y)}[F_{\phi}] - \log(  \mathbb{E}_{P(X)P(Y)}[e^{F_{\phi}}]) \\ 
    - 0.1 (\log(\mathbb{E}_{P(X)P(Y)}[e^{F_{\phi}}]))^2.
\end{multline}

\subsubsection{\textbf{Binary Cross Entropy}}

\begin{figure*}%[!t]
    \centering
    \includegraphics[width=1\linewidth]{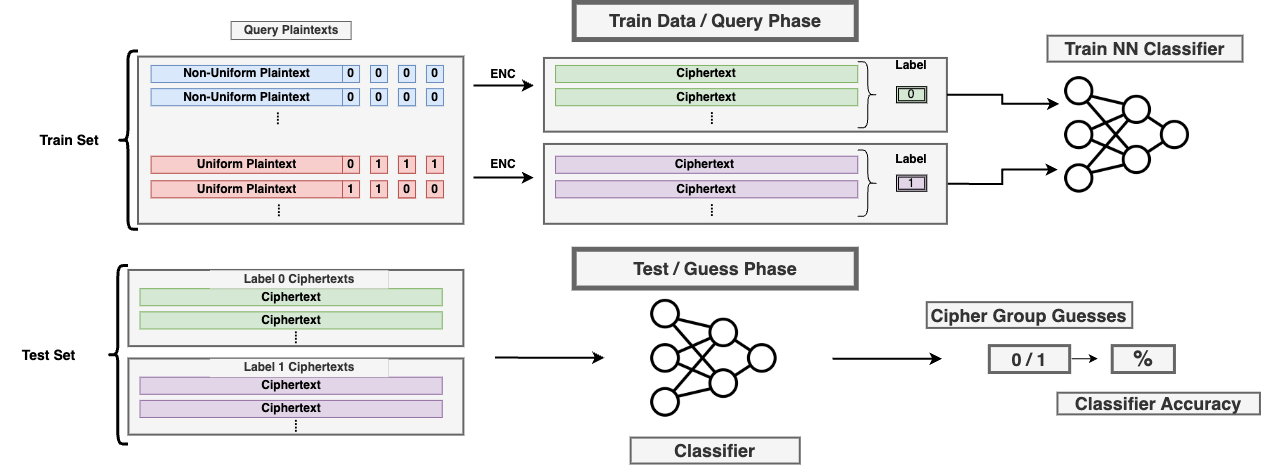}
    \caption{Framework for IND-CPA through Binary Classification Models}
    \label{fig:system_architecture}
\end{figure*}

Cross entropy is a widely used metric in the field of information theory to measure the distance between two probability distributions.

Binary Cross Entropy (BCE) is a specific form of cross-entropy loss tailored for binary classification tasks, where one can model a probability distribution, and predict between two classes\cite{goodfellow2016deep}. It measures the distance between two probability distributions, a true probability distribution and a predicted probability distribution. BCE quantifies the information lost when approximating the true distribution with the predicted distributions. We define the formula below and express $p$ as the true distribution, and $\hat{p}$ as the predicted distribution
\[\label{BCEeqn}
\text{BCE}(\hat{p}, p) = -\left[ \hat{p} \log(p) + (1 - \hat{p}) \log(1 - p) \right]. 
\]

In the context of machine learning, BCE is used as a loss function for binary classification, when tasked with classifying between two inputs. Since the function is differentiable, one can again optimize the function with SGD \cite{goodfellow2016deep}, and use a sigmoid function to output predictions in a binary format. When training a model with $N$ samples to represent a distribution, our objective is represented as 
\[\label{empiricalBCEeqn}
\text{BCE}(\hat{p}, p) = \frac{1}{N} \sum_{i=1}^{N} -\left[ p_i \cdot \log(\hat{p_i}) + (1 - p_i) \cdot \log(1 - \hat{p_i}) \right]. 
\]

This representation of BCE, also known as \emph{log loss} is fundamental in ML classification problems. In the following sections, we refer to this objective as BCE rather than log loss. Having defined all the relevant background and preliminaries, we now move on to the methods used in this work.

\section{Methodology}
\label{methodology}

%In this section, we describe all the steps and methods involved in performing our cryptanalysis. We present a cryptanalysis via mutual information neural estimation (Crypto-mine), which consists of estimation the MI between plaintext and ciphertext pairs. We also present our framework for IND-CPA through a \ac{BCE} loss classification model. We include specifics on preparing our datasets used, the architecture and training of our neural networks, and the specific steps and algorithms employed to estimate mutual information and perform binary classification. These methods will form the core of our cryptanalysis, enabling us to assess the security of a range of cryptosystems. \textcolor{blue}{introduction to methodology may need work}

This section presents our proposed method of cryptanalysis to evaluate the security level of a cryptosystem in terms of information theoretic metrics. We present the mutual information neural estimation, which consists of the estimation of the MI between plaintextand ciphertext pairs (Algorithm \ref{alg:MI_est}) as well as our framework for IND-CPA verification through a BCE loss classification model (Algorithm \ref{alg:BCEcpa}). The architecture of the proposed approach and procedures of the cryptanalysis are explained in detail. We also include the process of dataset creation and testing, specifics on the neural network structure, along with nuances of the algorithms employed to assess the security level of different cryptosystems.

\subsection{Dataset Preparation}\label{subsec:datasetprep}

To produce the datasets used in this paper, we first prepared two distinct groups of plaintexts: uniform plaintexts and non-uniform plaintexts for each cryptosystem. A non-uniform plaintext is defined as a plaintext consisting entirely of 0 bits, while a uniform plaintext is generated with randomly produced bits, with a different generation for each sample. MATLAB's built-in random number generator function was utilized to generate these uniform plaintexts. Next, we encrypted each plaintext batch with its respective cryptosystem, and finally labeled it for classification with either a ``0" or a ``1" for non-uniform plaintext and for uniform plaintext respectively. For all the encryption schemes tested, we re-use the same encryption key for all encryptions. All the dataset preparation was done in MATLAB, generating a total of 100,000 samples for training, and 20,000 for testing for every cryptosystem. There were an equal number of uniform/non-uniform plaintext samples -- 50,000 and 10,000 per category for the training and test set respectively.

\subsection{CRYTPO-MINE: Mutual Information Estimation on Cryptosystems}

By calculating the estimated mutual information leakage of an encryption scheme, an adversary is able to learn and potentially exploit an encryption scheme. We first prepare our plaintext and ciphertext pairs for our dataset (lines 1 and 2 of \ref{alg:MI_est}). We directly train the model through unsupervised learning, estimating the MI between the plaintext and ciphertext of an encryption scheme through calculating the joint and the marginals of the two distributions within each batch, measuring the distance with the KL-divergence (lines 3-7 of Algorithm \ref{alg:MI_est}), as defined in equation \ref{MINE_objective}.

\begin{algorithm}
\caption{MI Estimation for Cryptosystems}\label{alg:MI_est}
\begin{algorithmic}
\State 1: \textbf{Input} Plaintext \textbf{X} for \textbf{N} samples
\State 2: \textbf{Encrypt(x)} for ciphertext \textbf{Y} for \textbf{N} samples
\State 3: Initialize network parameters \textbf{$\theta$}
\State 4: \textbf{repeat}
\State 5: \hspace{0.25 cm} Find \textbf{I(X;Y)} between the sample set
\State 6: \hspace{0.25 cm} Compute SGD optimizing and updating \textbf{$\theta$}
\State 7: \textbf{until} convergence
\State 8: Iterate through and estimate MI of test set
\end{algorithmic}
\end{algorithm}

We provide our training dataset, which includes 100,000 samples of plaintexts and corresponding ciphertexts, to our neural network. Its architecture can be described as the following: the input nodes on the network vary depending on the length of the plaintext and ciphertext, and the intermediate nodes are set to either 100 nodes or 600 nodes with either 2 or 4 hidden layers respectively, depending on the dataset.  Running the experiments twice on two different sized networks gives us an idea on how the number of nodes in the neural network impact its abilities to capture dependencies between distributions in a cryptographic setting. We use ReLU nonlinearity \cite{goodfellow2016deep} for all layers and finally have one output node that results in the MI estimation for the batch. We set a learning rate of 1e-4, and the loss function as the DV representation of the KL-Divergence with a stabilizing term, defined in \ref{MINE_objective}. The model iterates through the data set for 1000 epochs with a batch size of 10,000 with SGD. Over the period of training, the estimate of MI will converge to a value, which will be high if the cryptosystem leaks MI between its plaintext and ciphertext. If the cryptosystem is leaking MI, our estimation will converge to a high MI. It is to be noted that depending on the strength of cryptosystem, this might happen quickly in a small number of epochs or take longer. However, if the cryptosystem is not leaking information, our results may still find slightly larger values due to overfitting of the correlations between plaintext and ciphertext. To mitigate the issue of overfitting, we include both the results from our test set and training set in our results. In a setting such as this one, where we perform a cryptanalysis, it is also interesting to see the training data as it indicates how successful the model is during training.

%\textcolor{blue}{Add how this helps to mitigate the issue.}

\subsection{Binary Cross Entropy for Classification}

Cross-entropy analysis is another useful metric for determining a cryptosystem's security \cite{cachin1997entropy}. One is able to calculate the cross entropy in a dataset between the plaintext and ciphertext by using the formula 

$$H(p_1, p_2) = -\sum_{x \in \mathcal{X}}p_1(x) log(p_2(x)),$$
 where $p_1$ represents the plaintext distribution, $p_2$ represents the ciphertext distribution and $\mathcal{X}$ is the set of samples. For a secure cryptosystem, we expect this value to be high, as it should take a lot of information to quantify the ciphertext from the plaintext or vice versa. While one may be able to gain valuable information about an encryption scheme's security, in this paper we use the case of BCE, defined in equation \ref{empiricalBCEeqn} to calculate the cross entropy between the distributions of encrypted ciphers originating from uniform plaintexts and encrypted ciphers originating from non-uniform plaintexts. We show that if we train a classification model to distinguish between different ciphers, the model is able to act as an adversary in a CPA setting.

\subsection{Modeling an IND-CPA game through Binary Classification}

The adversary in an IND-CPA experiment can be modeled through a binary classification model. This can be done with the following steps. We begin by queuing a polynomial number of plaintexts, both uniform and non-uniform (0’s), depicted in lines 1-3 in Algorithm \ref{alg:BCEcpa}. We label the corresponding non-uniform ciphertexts with 0, and uniform ciphertexts with 1 (lines 5 and 6 in Algorithm \ref{alg:BCEcpa}), and we use supervised training to train the model inserting each ciphertext with its corresponding label (lines 8-12 in Algorithm \ref{alg:BCEcpa}). Querying these plaintexts aligns with the second step in our definition of the IND-CPA experiment.

We then can send one uniform and one non uniform plaintext for the guessing phase of the IND-CPA experiment, which corresponds to the third step in Definition \ref{INDCPA_defn}. Finally, we test the corresponding ciphertext that was sent back with our classification model (lines 13 and 14 in Algorithm \ref{alg:BCEcpa}), consistent with the final step of Definition \ref{INDCPA_defn}. We repeat this guessing process with a large batch of ciphers for our model to gauge the model’s accuracy. As shown in Section \ref{results}, the model proves to be most effective when the cross entropy disparity between non-uniform and uniform plaintexts can be exploited.

\begin{algorithm}
\caption{IND-CPA BCE Classification Adversary}\label{alg:BCEcpa}
\begin{algorithmic}
\State 1: \textbf{Input} Non-uniform plaintext set \textbf{$X_0$} to challenger
\State 2: \textbf{Input} Uniform plaintext set \textbf{$X_1$} to challenger
\State 3: \textbf{Challenger} outputs ciphertext sets \textbf{$Y_0$}, \textbf{$Y_1$} from \textbf{$X_0$}, \textbf{$X_1$} 
\State 4: Label ciphertexts \textbf{$Y_0$} with ``0" for all in set
\State 5: Label ciphertexts \textbf{$Y_1$} with ``1" for all in set
\State 6: \textbf{Form dataset $Y$} from \textbf{$Y_0$} and \textbf{$Y_1$}
\State 7: Initialize network parameters \textbf{$\theta$}, 
\State \hspace{0.25 cm} predicted distribution $\hat{Y}$
\State 8:  \textbf{repeat}
\State 9: \hspace{0.25 cm} Find \textbf{BCE($\hat{Y};Y)$} between the sample set
\State 10: \hspace{0.25 cm} Compute SGD optimizing and updating \textbf{$\theta$}
\State 11: \textbf{until} convergence
\State 12: Send one uniform, one non-uniform challenge plaintexts to challenger
\State 13: Have \textbf{$\theta$} classify challenge cipher, output guess
\end{algorithmic}
\end{algorithm}

We note that our BCE indistinguishability classifier works with any two different groups of plaintexts-- for example a plaintext group with all 0 bits and a plaintext group of all 1 bits. We select the two groups of uniform and non-uniform plaintexts because the distance between each group's distribution is the greatest. In certain cases, this also results in the greatest distance (if any) between the distributions of each groups ciphertexts, which is optimal for our cross entropy classifier since it measures the distance between the two distributions.

For our classification experiments, as elaborated in Section \ref{subsec:datasetprep}, we prepare a dataset of plaintexts and ciphertexts of 100,000 samples, containing both uniform and non-uniform plaintexts, with their respective ciphertext pairs. We also prepare a test set with 20,000 pairs. Our neural network either contains two intermediate layers of 100 nodes each, or four intermediate layers of 600 nodes each. We label each ciphertext with its respective label, and train our BCE classification model with supervised learning \cite{goodfellow2016deep}. We set our loss function to the BCE redefined for samples, and iterate through the data set for 1000 epochs with a batch size of 10,000. We calculate and optimize the model's predicted distribution using SGD, and use ReLU non-linearity for all the layers except the output layer, where sigmoid is used. If the calculated cross entropy is near 0, the predicted distribution is close to estimating the true distribution, and our classification model can either break the cryptosystem in an IND-CPA setting or has overfitted the data. Again, to prevent overfitting, we use the separate test set prepared to gauge the model's accuracy. All of our experiments were trained on a NVIDIA 2080 TI GPU.

The two methods presented yield different metrics through different methods-- i.e. MI estimation and binary classification, through unsupervised and supervised learning respectively. However, they have two fundamental similarities. Firstly, both algorithms are deployed in a known/chosen plaintext setting, and secondly, both algorithms use information theoretic metrics to estimate probability distributions based on the samples, which are then used to calculate the loss and optimize the models. Both these approaches provide an empirical analysis of two different aspects of security and thus useful practical verification of the robustness of the cryptosystem.

\section{Experimental Results}
\label{results}

With our definitions, methods, and algorithms defined, we now present our experimental results. We first verify our methods on several basic cryptosystems. Following empirical verification of effectiveness, we employ the methods to analyze widely recognized cryptosystems \cite{standard1999data, daemen1999aes, rivest1978method} and network coding-based cryptosystems \cite{cohen2021network}.

For our baselines, we test the following schemes: no encryption, a constant XOR key, and a one time pad \cite{shannon1945mathematical}. We then test the following standardized schemes: Rivest–Shamir–Adleman (RSA) Encryption \cite{rivest1978method}, Advanced Encryption Standard (AES-128) \cite{daemen1999aes}, and Data Encryption Standard (DES) \cite{standard1999data}. We test two modes for each of these schemes, one which is deterministic such as AES Electronic Code Book (ECB), Plain RSA, and Standard DES, and one that is probabilistic and either CPA or Indistinguishability under Chosen Ciphertext Attack (CCA) secure, such as AES Counter (AES CTR), a non-deterministic DES padded with random bits before encryption, and Optimal Asymmetric Encryption Padding (OAEP) RSA \cite{bellare1995optimal}. With results from these encryption schemes, one is then able to perform the same cryptanalysis on network coding based cryptosystems \cite{cohen2021network, cohen2022partial, woo2023cermet}, and compare the results with the standardized RSA, DES, and AES, as well as our baselines.

We also consider other factors such as computational power, and how a larger network with more nodes impacts learning capacity. Therefore, excluding the simple baselines, we perform our experiments twice as mentioned in Section \ref{methodology}. Once on a smaller neural network which contains two intermediate layers of 100 nodes, and once on a larger neural network which contains four intermediate layers of 600 nodes. We use the same dataset and test set for both MI estimation and our BCE classifier indistinguishability experiment.

\begin{table}[H]
\centering
\begin{tabular}{|>{\centering\arraybackslash}m{2.5cm}|>{\centering\arraybackslash}m{2cm}|>{\centering\arraybackslash}m{2cm}|}
\hline
\textbf{Encryption} & \textbf{Test MI (nats)} & \textbf{IND-CPA accuracy} \\
\hline
No Encryption & 9.17 & 100\% \\
\hline
One time pad & 0.0092 & 50.36\% \\
\hline
Constant Key XOR & 5.16 & 100\% \\
\hline
\end{tabular}
\caption{Test Set Baselines Results}
\label{table:Baselines}
\end{table}

\subsection{Baselines}

\begin{figure}[!t]
  \centering
  \includegraphics[width=\linewidth]{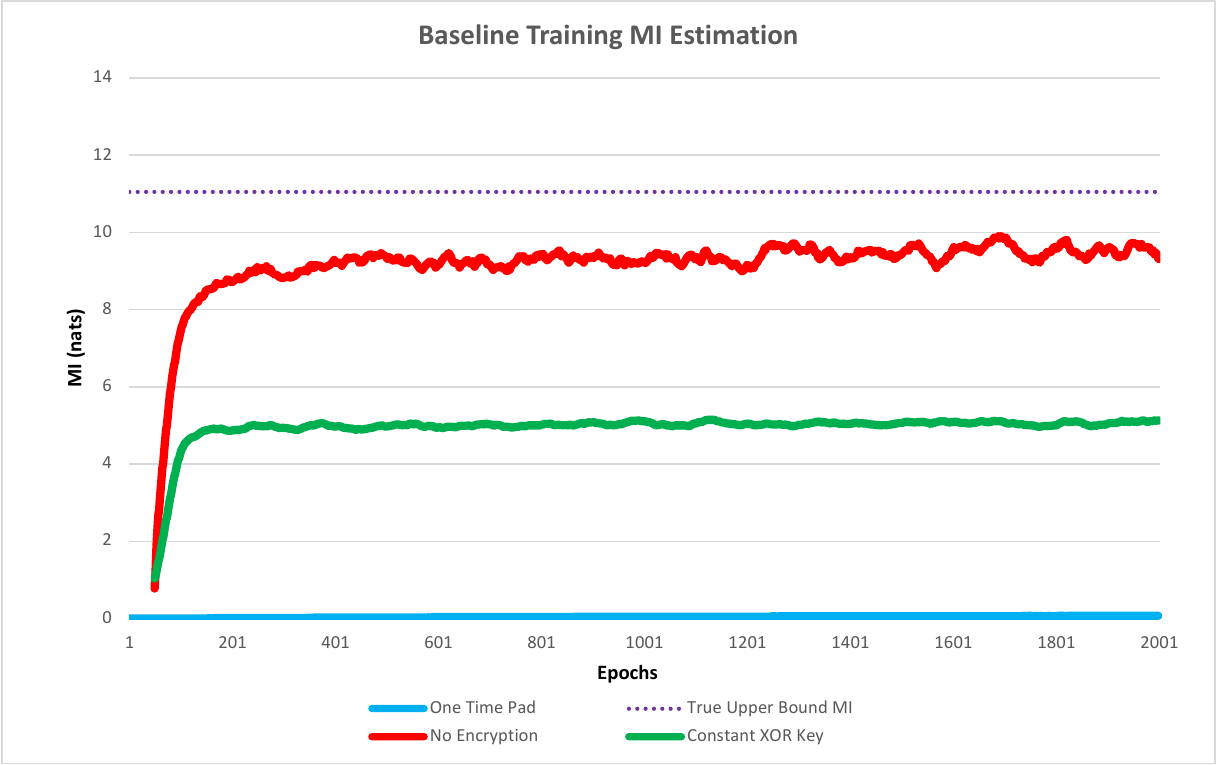}
  \caption{MI Estimation Training Results for Baselines.}
  \label{fig:baselines_mi_training}
\end{figure}

We begin by experimenting on baseline encryption settings, such as a constant XOR key, a one time pad, and no encryption. We can get the true MI of no encryption, by calculating $I(X;Y) = H(X) - H(X|Y), H(X|Y) = 0$ since $ X = Y$ and $H(X) = 11.1$ nats $ = I(X;Y)$. We can use this calculated MI as an upper bound MI for any cryptosystem, and we see that our estimation illustrated in the training data (Figure \ref{fig:baselines_mi_training}) and test set (Table \ref{table:Baselines}) of 9.6 and 9.17 nats respectively, performs well. Also, we expect any competent adversary to classify all the plaintext and ciphertexts correctly in an IND-CPA setting, and we do achieve 100\% accuracy for 20,000 samples in our test set when testing our BCE classifier with no encryption. Next, on the other side of the security spectrum, we test a one time pad approach which is expected to provide perfect secrecy. We expect an estimation of zero MI and that our classifier cannot do any better than randomly guessing, giving us 50\% classification accuracy. Both expectations are satisfied upon experimentation as demonstrated in \ref{table:Baselines} and \ref{fig:baselines_mi_training}. A negligible MI estimate of 0.0092 nats is calculated, since $\phi$ in our MI estimator attempts to approximate a function given our datasets. Moreover, we correctly classify at an accuracy of around 50\%, as anticipated for a perfectly secure scheme. Lastly, we test a simple single key XOR where a singular key is reused for every encryption. The single key XOR results in an estimated MI of 5.16 and classification accuracy in a IND-CPA experiment setting of 100\%, both consistent with expectations for such a simple encryption scheme. These baseline experiments provide empirical validation in our approaches and we extend the analysis to more widely used cryptosystems. %Lastly, we test classical ciphers non-secure ciphers such as a Vingere cipher and Block cipher. These schemes leak less MI than our insecure baselines, but more than secure cryptosystems we test in the next section, and can also be classified with 100\% accuracy.

\begin{table}[H]
\centering
\begin{tabular}{|>{\centering\arraybackslash}m{2.5cm}|>{\centering\arraybackslash}m{2cm}|>{\centering\arraybackslash}m{2cm}|}
\hline
\textbf{Cryptosystem} & \textbf{Small NN Test MI} & \textbf{Big NN Test MI }\\
\hline
DES          & 0.733       & 1.916     \\
\hline
DES (Non-deterministic)   & 0.138       & 1.795     \\
\hline
AES ECB      & 0.0621      & 0.7068    \\
\hline
AES CTR      & 0.0335      & 0.263     \\
\hline
Plain RSA    & 0.7285      & 1.481     \\
\hline
Padded RSA   & 0.0421      & 1.342     \\
\hline
\end{tabular}
\caption{Test Set MI Estimates for Cryptosystems}
\label{table:MI_TEST_POPULAR_CRYPTO}
\end{table}

\subsection{Mutual Information Estimation Results}

\begin{figure}[!t]
  \centering
  \includegraphics[width=\linewidth]{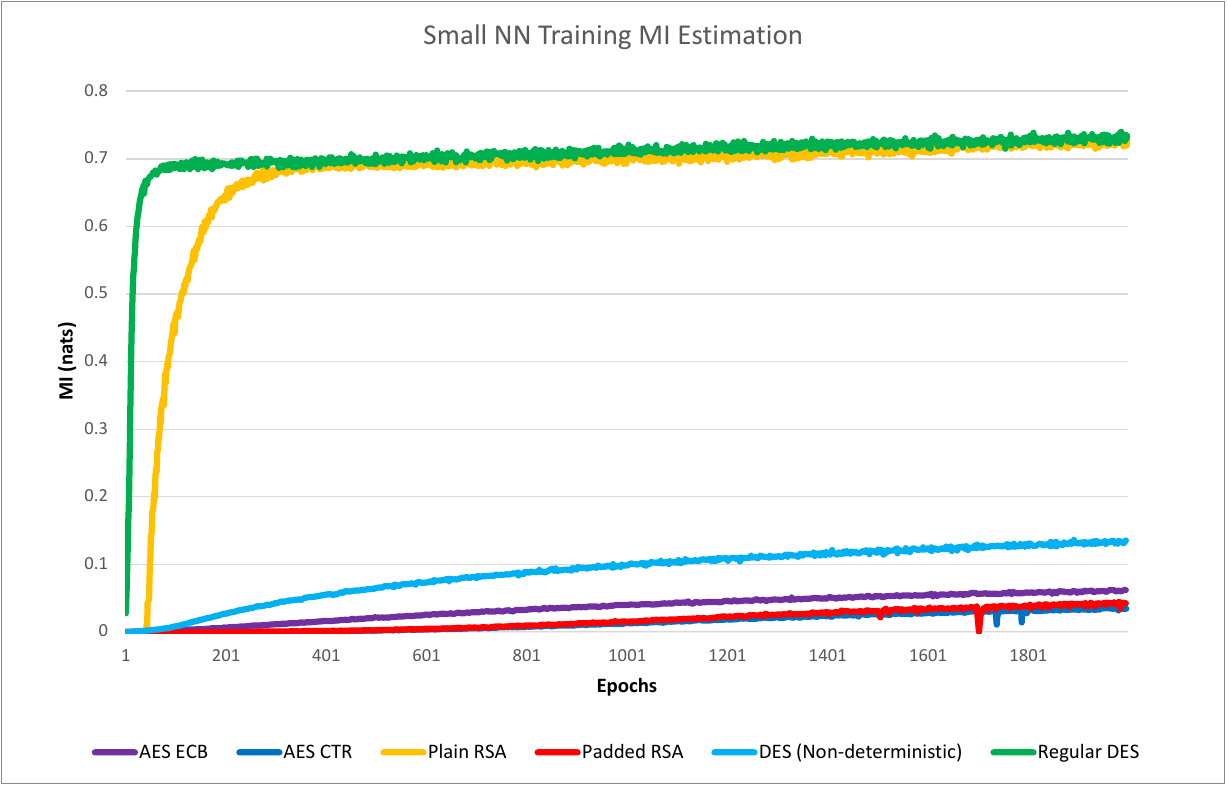}
  \caption{MI Estimation Training Results for Several Cryptosystems on a Smaller Neural Network.}
  \label{fig:MI_smaller_nn_graph}
\end{figure}

\begin{figure}[!t]
  \centering
  \includegraphics[width=\linewidth]{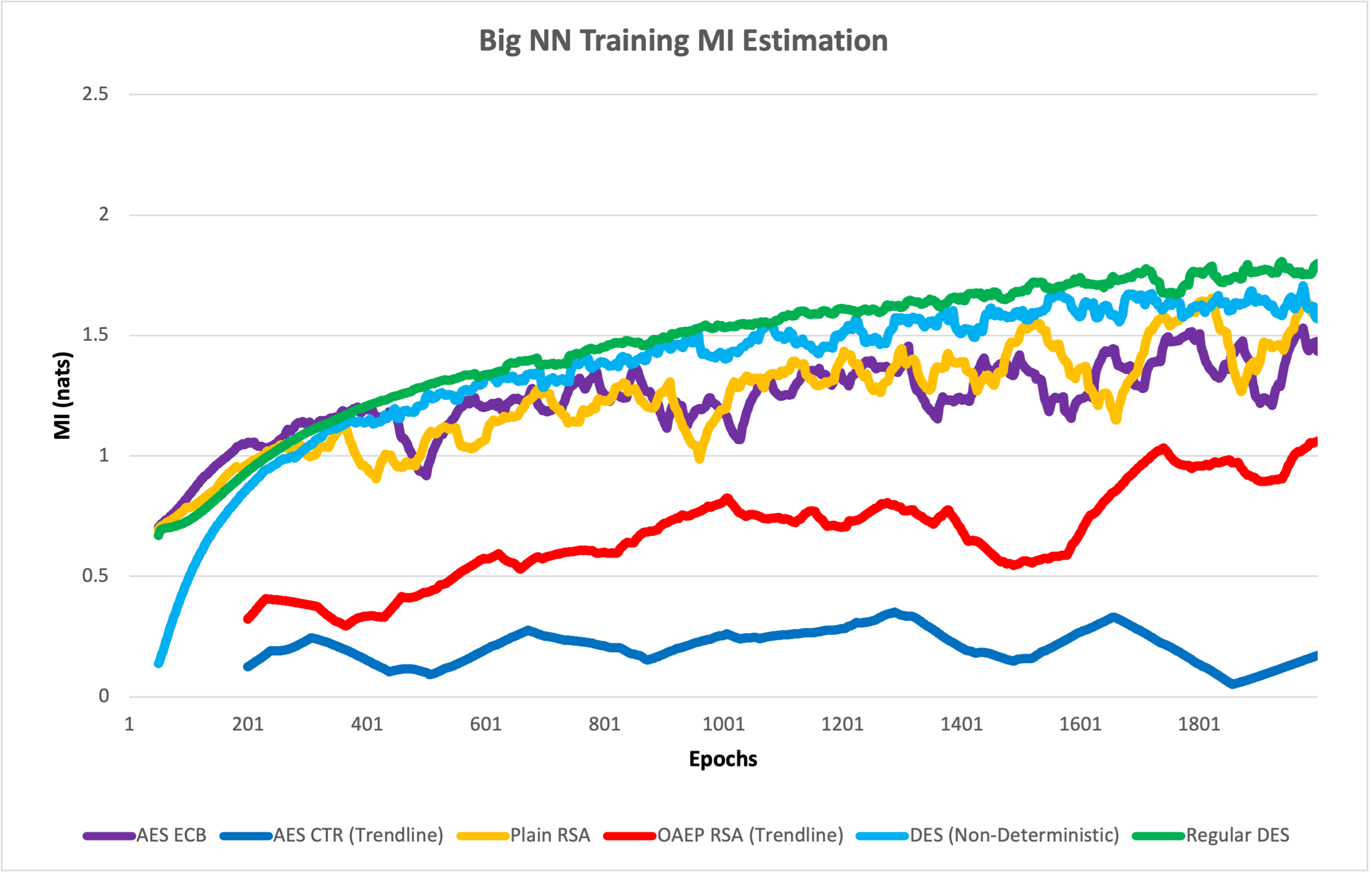}
  \caption{MI Estimation Training Results for Several Cryptosystems on a Larger Neural Network.}
  \label{fig:BIG_NN_MI_TRAINING}
\end{figure}

With our baselines demonstrating the effectiveness of our cryptanalysis techniques, we test our MI estimator on several widely-used encryption schemes including modes of RSA, AES, and DES. Figure \ref{fig:BIG_NN_MI_TRAINING} presents the training data for our MI estimation of different cryptosystems with our larger neural network of 600 nodes and four intermediate layers (Big NN). Figure \ref{fig:MI_smaller_nn_graph} represents the training data for our MI estimations with the smaller network with two intermediate 100 node layers (Small NN). The MI estimated from our test sets for both networks are provided in Table \ref{table:MI_TEST_POPULAR_CRYPTO}. One can see that with both the smaller and larger networks the results are consistent, for both training and testing sets, with DES leaking the most MI, followed by RSA, and AES leaking the least amount of MI. The larger network is able to calculate a larger MI leakage for the cryptosystems, since the neural network $\theta$ has 2200 (or 12x) more internal nodes to create a function ($\phi$) that maximizes the MI between the plaintext and ciphertext. The larger network also has more variance during training, due to being a more complex model and having more parameters. When analyzing the results, we see that deterministic schemes such as AES ECB, DES, and Plain RSA leak more MI than their probabilistic counterparts, AES CTR, Padded DES, and OAEP RSA. This is due to $\phi$'s network being able to map a deterministic function between the plaintexts and ciphertexts for each encryption scheme, whereas it is unable to do so for the probabilistic encryption schemes. 

It is also noteworthy that Crypto-mine's MI estimation for its test sets reflects the MI estimations during training. If the model was overfit to the data but unable to truly detect any MI leakage for a cryptosystem, the test MI would be close to zero. However, we see that this is not the case for any of our trials -- our test MI is consistent with the maximum MI achieved during training for the most part. These results help us conclude that an accurate MI estimator with large computing power may potentially provide useful information to adversaries and attackers about encryption schemes in a known plaintext attack setting.

\subsection{Binary Cross Entropy Classification for Indistinguishability Analysis Results}

\begin{table}[h]
\centering
\begin{tabular}{|>{\centering\arraybackslash}m{2.5cm}|>{\centering\arraybackslash}m{2cm}|>{\centering\arraybackslash}m{2cm}|}
\hline
\textbf{Cryptosystem} & \textbf{Small NN IND-CPA Classification Accuracy} & \textbf{Big NN IND-CPA Classification Accuracy }\\
\hline
DES          & 100\%       & 100\%     \\
\hline
DES (Non-deterministic)   & 50.32\%        & 50.03\%     \\
\hline
AES ECB      & 100\%     & 100\%    \\
\hline
AES CTR      & 49.98\%      & 50.51\%  \\
\hline
AES CTR (Reduced Counter Field)     & 75.25\%      & 98.92\%     \\
\hline
Plain RSA    & 100\%      & 100\%     \\
\hline
Padded RSA   & 50.2\%      & 49.45\%     \\
\hline
Padded RSA (Re-Used Padding)  & 61.23\%    & 99.5\%     \\
\hline
\end{tabular}
\caption{IND-CPA Classification Accuracy for Cryptosystems}
\label{table:INDCPA-crypto}
\end{table}

Following our MI leakage analysis for modes of RSA, AES, and DES, we now present our results for our IND-CPA tests using BCE classifier model. One should not expect a BCE classification model to break any CPA/CCA secure schemes. Therefore when testing the CPA/CCA secure versions of these encryption schemes, we also test cases where a cryptosystem's security may be compromised by an error, such as resetting the counter for AES CTR during encryption, or re-using padding seed and hash outputs for (now non-optimal) asymmetric encryption padding for RSA. These test cases are consistent with our algorithm's auditing abilities, where such errors may occur over large datasets.

The results are shown in Table \ref{table:INDCPA-crypto}. As depicted, our classifier performs extremely well on deterministic versions of encryption, such as AES ECB, DES, and Plain RSA, with classification accuracies of 100\% for each encryption mode. This is expected, as these deterministic encryption schemes are known to not be IND-CPA secure. However, the classifier struggles to correctly classify CPA and CCA secure encryption schemes, such as AES-CTR, non-deterministic DES, and Padded RSA achieving classification accuracies with negligible variance from 50\%. We re-emphasize that we do not expect an entropy based classifier to break CPA secure encryption schemes.

With this in mind, we also test our model with these encryption schemes where a user may not be encrypting the data properly, which would allow an adversary to potentially break these schemes. Consider the cases where a user reuses padding while encoding and encrypting with OAEP RSA, or resets the counter in AES CTR to zero while using a constant key and initialization vector. We find that our classification model is able to exploit the compromised security in these cases.

For the counter experiment, we set our counter field to reset at 100,000 when generating our samples, which lead to a 99\% classification accuracy for the larger neural network under our IND-CPA game conditions. This is a significant increase, as the model goes from no better than randomly guessing, achieving an accuracy of 50\% with negligible variance to only classifying 204 out of the 20,000 ciphers, to a classification accuracy as high as 99\% once the counter is overrun.

We also test the CCA secure OAEP RSA where we re-use the padding seed and hash when padding our message during encoding every 100,000 samples such that samples in the training and test sets have the same padding. Here, we again achieve a similar improvement, with an accuracy of 99.5\%. In these cases where a CPA secure scheme is modified leading to compromised security, the classification model performs extremely well, identifying the vulnerability in the modified schemes with our IND-CPA framework. 

The results demonstrate the efficacy of our BCE indistinguishability classifier framework in compromising non-CPA secure schemes. The model performs exceptionally well in distinguishing between different ciphertexts for non-CPA encryption methods. Moreover, it succeeds in exploiting vulnerabilities in CPA-secure schemes when we introduce slight modifications in the encryption process. Our framework coupled with our experimental results highlights the versatility and potential of ML classification for chosen plaintext attacks.

\subsection{Cryptanalysis on Network Coding Based Cryptosystems}\label{cryptanalysis_HUNCC}

\begin{figure}[!t]
  \centering
  \includegraphics[width=\linewidth]{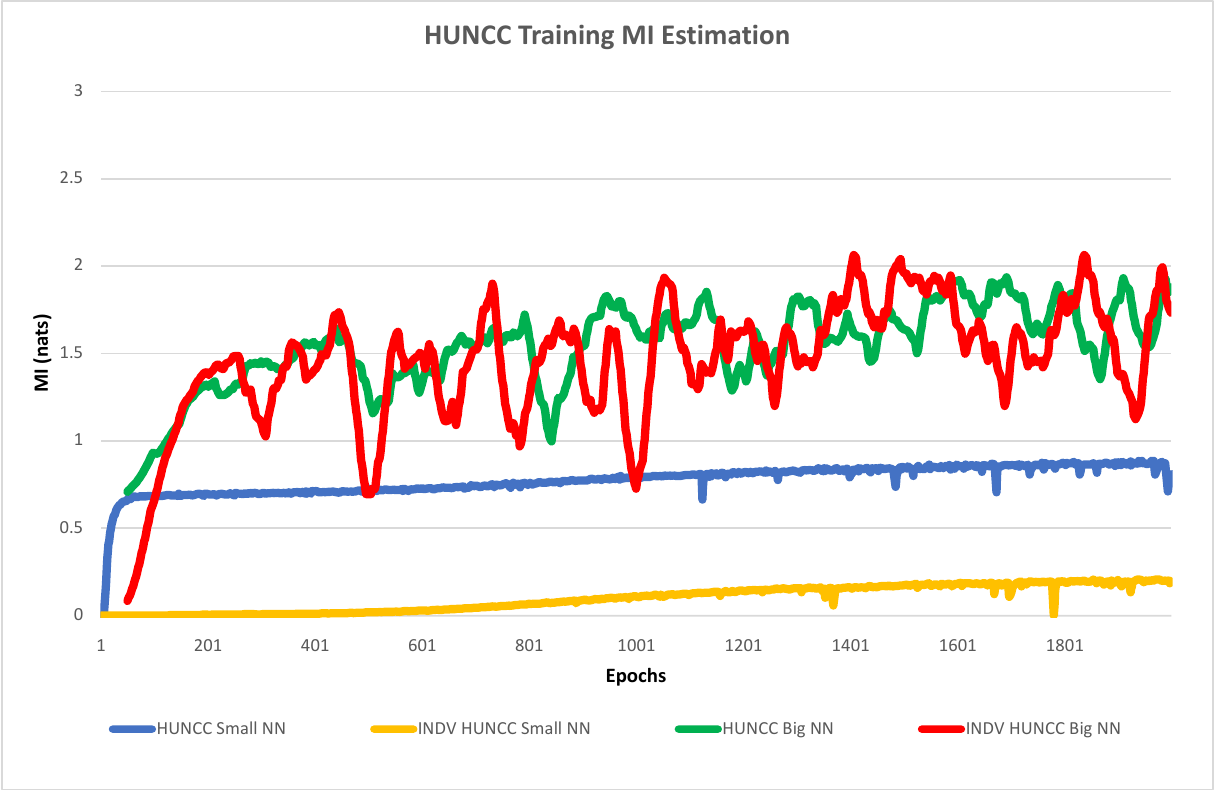}
  \caption{MI Estimation Training Results for HUNCC.}
  \label{fig:HUNCC_training}
\end{figure}

\begin{table*}[t]
\centering
\begin{tabular}{|c|c|c|c|c|}
\hline
Cryptosystem & Small NN Test MI & Small NN IND-CPA & Big NN Test MI & Big NN IND-CPA \\ \hline
HUNCC & 0.822 nats & 100\% & 2.137 nats & 100\% \\ \hline
Individual Secrecy HUNCC & 0.201 nats & 49.09\% & 1.782 nats & 50.01\% \\ \hline
\end{tabular}
\caption{HUNCC Cryptanalysis Results}
\label{table:HUNCC_table}
\end{table*}

With our results and analysis of traditional and popularized encryption schemes, we extend our cryptanalysis beyond these schemes to network coding-based encryption schemes, specifically the Hybrid Universal Network Coding Cryptosystem (HUNCC), presented in \cite{cohen2021network}. HUNCC utilizes code-based cryptography, allowing one to achieve security guarantees in a network setting while partially encrypting coded messages to transmit data at a much more efficient rate. HUNCC also provides an energy efficient alternative to expensive post-quantum cryptosystems such as McEliece, where the costly encryption can be used in a smaller portion of the data with precoding \cite{woo2023cermet}. 

Particularly, HUNCC ensures individual security for the messages and is proven to be individually IND-CCA1 secure if the underlying cryptosystem is IND-CCA1 secure \cite{cohen2022partial}. The IND-CCA1 security guarantee proved in \cite[theorem 2]{cohen2022partial}, also implies individual IND-CPA security, defined in \ref{IndividualINDCPA_DEFN} if the underlying cryptosystem is IND-CPA secure. Individual IND-CPA provides a computational analogue to individual security, as it provides a setting where an adversary is only able to choose a message in a singular channel, and guarantees that they are only able to learn a negligible amount of information from that singular message channel. %This is done by first linearly coding the data and then encrypting a portion of it. 
More specifics on HUNCC's algorithm and individual indistinguishability can be found in Appendix \ref{HUNCC_indepth}.

% \textcolor{purple}{HUNCC algorithm in appendix} \textcolor{blue}{ add more background to HUNCC. Also be sure to mention the notion of individual security if we do not in the introduction, Individual IND-CPA and HUNCC's algorithm will be defined in the appendix}

\begin{figure}[h]
  \centering
  \includegraphics[width=\linewidth]{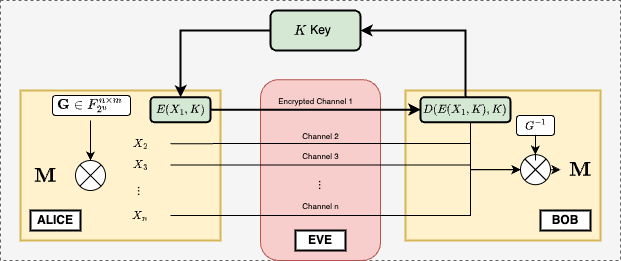}
  \caption{HUNCC Overview. A detailed algorithm and description can be found in Appendix \ref{HUNCC_indepth}}
  \label{fig:HUNCC_OVERVIEW}
\end{figure}

For the analysis of the HUNCC scheme in this work, we consider HUNCC defined in $GF(2^8)$, with 8 outgoing channels that each contain 16 bytes. We then encrypt one of these 8 channels using AES CTR. This results in a total plaintext and ciphertext length of 128 bytes each. For these experiments, we use the same 8 by 8 generator matrix for all our samples. For our experiments testing for individual secrecy, the dataset is modified as follows: 7 of the 8 channels in the plaintexts are uniformly chosen for all samples, while the last channel can be chosen to either be non-uniform or uniform. This representation of the CPA/CCA experiments can be found in \cite{cohen2022partial}, enabling one to see if there is information leakage individually by channel, and represents a computational analogue to the concept of individual security. More on this modified CPA experiment can be found in the appendix.

The repeated use of the encoding matrix leads to us not being able to pass the BCE indistinguishability test. When we test for HUNCC for individual secrecy, allowing for only one of the channels to be chosen in an Individual IND-CPA, detailed in the appendix and \cite{cohen2022partial}, our proposed BCE classification framework can no longer correctly classify the samples. The outcome of HUNCC being broken under indistinguishability, but not individual indistinguishability is anticipated and verifies the individual secrecy guarantee of HUNCC, proved in \cite[theorem 2]{cohen2021network}.

The BCE indistinguishability and MI estimation results are presented in Table \ref{table:HUNCC_table} and Figure \ref{fig:HUNCC_training}. We can see that the MI leakage for HUNCC for both the big and small neural network is comparable to those of the standardized encryption schemes in the \ref{table:MI_TEST_POPULAR_CRYPTO}, leaking a maximum MI of around 2 nats, even with the same deterministic coding matrix for all the samples. The similar MI leakage from HUNCC when normally tested and when tested for individual secrecy indicates the fact that the larger neural network is able to map a similar function both in a normal setting and the setting where we test for individual secrecy. It is also worth noting how the smaller neural network is able to detect more MI leakage for HUNCC when we use the fully non-uniform standard dataset leading to more determinism between plaintext and ciphertext pairs. Compared to dataset testing for individual secrecy, we see much less determinism due to only one of the eight channels being chosen as non-uniform, which leads to more stochasticity in our corresponding ciphertexts. In contrast, the big neural network is able to map a similar function to $\phi$ for both the standard dataset and the dataset for individual security, given it has 12x the internal nodes to do so. A more in-depth study on HUNCC's input uniformity correlating to MI leakage can be found in \cite{kim2024crypto}, and a version of HUNCC for non-uniform plaintexts (NU-HUNCC) can be found in \cite{10619109}. Extending our analysis to the NU-HUNCC can be an interesting future work.

% \section{\textcolor{yellow}{usecases}}

% do we want this section or is it ok to include them like I did in the discussion.

\section{Discussion}
\label{discussion}

% \textcolor{yellow}{Broken into the following Chronologically: Further discussions on results, Use Cases and Novelty, Information measures and cryptography, MINE, Classification models, ML for cryptanalysis, HUNCC,Future work}

The proposed methodology and results offer an interesting perspective on known/chosen plaintext attacks in the field of cryptography, and a novel application of machine learning models. Our results reflect promising capabilities of both the classifier and MI estimator, with the classifier being able to identify all of our non-CPA secure schemes with high accuracy, while also confirming CPA secure schemes to be secure. Our results also indicate that a smaller sized neural network suffices with finding vulnerabilities within our encryption schemes, while we find usefulness in a larger network size, when we would like to fully break a scheme. Moreover, our MI estimator first demonstrates effectiveness by accurately measuring simple encryption scheme's MI leakage, followed by measure the MI leakage of popularized encryption schemes. Being able to estimate leakage in correlation between an plaintext and ciphertext for encrpytion schemes has been proven to be useful. While both of the algorithms presented yield different metrics, they both use information theoretic metrics to estimate distributions and provide useful unique information about a cryptosystem through empirical bounds obtained. Moreover, both are easily implementable with a singular dataset. These frameworks are universal for any proposed cryptosystem or communication setting and can readily be applied in real-world use cases. 

One of the most promising applications of this approach is to perform security auditing. If a user wants to send sensitive data in the open, they might be using a particular encryption scheme for a long time. It is necessary to ensure that the continuous use of the encryption scheme does not deteriorate the robustness of the scheme. Our approaches provide empirical analysis on the efficacy of cryptosystems as well as potential mistakes that could lead to leakage of information or exposure to indistinguishability. Use cases for auditing encrypted data include data centers transferring large amounts of encrypted data, healthcare providers sharing encrypted sensitive data in the open, and companies handling lots of sensitive information.

Another use case is the applied testing of proposed encryption schemes, such as HUNCC. While getting promising results from our frameworks doesn't formally guarantee security, the empirical bound on security constraints provided by our frameworks for cryptographic protocols can be useful in practical settings. Particularly, for code-based cryptographic approaches like HUNCC, the leakage of MI between the inputs deteriorates the security levels. Even the non-uniformity in the input distribution can also impact the security level. Theoretically, HUNCC requires the inputs to be uniformly distributed. However, for practical purposes, the level of randomness in the input distribution can be smaller. With our approach, we could provide more insights into this level of uniformity that can be accepted in the input distribution and still does not leak a significant amount of MI in the coded, partially encrypted output.%As data encryption and decryption becomes more and more expensive to combat quantum computing power, an emerging area of energy efficient cryptography is being explored \cite{goodman2001energy,le2009energy,woo2024leveraging}. Our algorithms and frameworks can be useful aids in the applied testing proposed schemes.

%\textcolor{blue}{talk about HUNCC for a couple of sentences here}

% \subsubsection{Information Measures and Cryptography}
% \textcolor{blue}{talk about Information measures and crypto}

\subsubsection{Mutual Neural Information Estimation } MINE has represented a significant improvement in estimating mutual information from high dimensional data successfully, being able to estimate the joint and marginal distributions of variables. The use of neural networks allows MINE to  capture complex, non-linear dependencies between variables through unsupervised learning. Although issues of stability and variance arise, it has also proved useful in several applications such as learning representations \cite{hjelm2018learning}, data shaping \cite{esfahanizadeh2023infoshape}, and generative models. Our application to the field of cryptography presents a novel method of calculating this measure in a known plaintext setting. %\textcolor{red}{more about MINE}.

\subsubsection{Classifiers } Machine learning classifiers are essential tools for predictive modeling, encompassing a range of algorithms tailored for various types of data and applications. Neural networks, particularly deep learning models, have been proven to excel in handling underlying patterns and large datasets, making them suitable for complex tasks. Applications of these classifiers highlight their influence and versatility, spanning across fields including healthcare, finance, and biology. BCE loss classifiers are one of the most fundamental and popular classifiers in the field. Our simple yet elegant approach of being able to model an IND-CPA experiment with a prepared dataset, is intended to serve as a foundation for ML models performing a cryptanalysis while following one of cryptography's most prominent and foundational security experiments. Deep learning models' pattern recognition and classification capabilities, as well as their ability to learn underlying dependencies between distributions, make them potentially an optimal aid to adversaries in cryptographic settings.

\subsubsection{ML for Cryptanalysis} It is interesting to note all the different presented ML approaches which can be utilized for cryptanalysis. Both fields share the goal of creating a function based on a set of inputs, and though historically different approaches are taken to achieve this goal, the use of ML approaches can potentially have a revolutionary impact. Within recent years, more and more attacks utilizing ML models have been presented, both for classical cryptography and post quantum cryptography (PQC). Attacks on classical encryption schemes, such as the encryption modes studied in this work, have been successful on reduced round block ciphers \cite{gohr2019improving}, and substitution ciphers \cite{kambhatla2018decipherment}. Notable PQC attacks utilizing machine learning algorithms to perform a side-channel analysis of Kyber \cite{dubrova2023breaking}, an encryption scheme standardized by NIST \cite{NIST-PQC}, and using transformers to solve size-reduced learning with errors problems \cite{wenger2022salsa}.

\section{Conclusion}
\label{conclusion}
%[Need to extend this]
In this paper, we present two novel applications of machine learning algorithms that allow one to perform cryptanalysis on any encryption scheme of their choice. While the mutual information estimation using neural networks evaluates the encryption scheme against conditions of perfect secrecy, the binary cross entropy classifier models an indistinguishability under chosen plaintext experiment that is commonly used in modern cryptography. These empirical evaluations of the cryptosystems using novel techniques demonstrate the potential vulnerabilities in widely used encryption methods and provide insights into their security properties. Our analysis included the evaluation of several standardized cryptosystems and their different modes, followed by cryptanalysis of a network coding-based cryptosystem showcasing the versatility and effectiveness of our proposed machine learning-based methods. The results highlight potential vulnerabilities in these cryptosystems and underscore the importance of using advanced analytical techniques to ensure robust security. The universality of the frameworks presented leads to several potential applications and use cases. Moreover, the information theoretic metrics and calculations from the models can provide useful insight for an adversary or a user wishing to perform a security audit. 

By leveraging advanced machine learning approaches, we aim to offer a deeper understanding of cryptographic strength and foster the development of more robust encryption protocols. Our study reinforces the effectiveness of machine learning approaches in cryptanalysis and presents a new direction of work where the two fields are intertwined. Future work based on our proposed frameworks includes designing more specialized ML models and frameworks to target specific secure encryption schemes and attacking schemes with the calculated MI leakage.

\appendix

\subsection{Hybrid Universal Network Coding Cryptosystem Algorithm }\label{HUNCC_indepth}

\begin{algorithm}
  \caption{HUNCC Scheme}
  \label{alg:hybrid_scheme}
  \begin{algorithmic}[1]
    \Statex\textbf{Input:}  At Alice, \( n \) data splits/channels \( [M_1;\ldots;M_{n}] \in \mathbb{F}^{n}_{q^m} \)
    \Statex\textbf{\underline{Encoding at Alice}:}
    \State \textbf{Stage 1:} MRD secrecy encoding of each split
    \For{each split \( i \) in \( [M^{(1)}, \ldots, M^{(N)}] \)}
        \State \( X^{(i)} =  \textbf{G} M^{(i)} \)
    \EndFor
    \State \textbf{Stage 2:} Encrypt selected splits
    \For{each chosen split \( 1 \leq i \leq c \)}
        \State \( \ddot{\mathbf{b}}_{i} \leftarrow [X_{i}^{(1)},\ldots, X_{i}^{(N)}] \)
        \State \( \mathbf{y}_i = \mathrm{Enc}(\ddot{\mathbf{b}}_i, p_i) \)
    \EndFor
    \Statex\textbf{\underline{Transmission}:}
    \For{each split \( 1 \leq i \leq n \)}
        \State Transmit \( \mathbf{y}_i \)
    \EndFor
    \Statex\textbf{\underline{Decoding at Bob}:}
    \State \textbf{Stage 1:} Decrypt received data
    \For{each encrypted split \( 1 \leq i \leq c \)}
        \State \( [X_{i}^{(1)},\ldots, X_{i}^{(N)}] \leftarrow \mathrm{Dec}(\mathbf{y}_i, s_i) \)
    \EndFor
    \State \textbf{Stage 2:} Decode each split
    \For{each split \( i \) in \( [X^{(1)}, \ldots, X^{(N)}] \)}
        \State \( (M_{1}^{i};\ldots;M_{n}^{i}) = \textbf{H} X^{i} \)
    \EndFor
    \Statex\textbf{Output:} At Bob, \( [M_1;\ldots;M_{n}] \in \mathbb{F}^{n}_{q^m} \)
  \end{algorithmic}
\end{algorithm}

In Section \ref{cryptanalysis_HUNCC}, we analyzed the hybrid cryptosystem HUNCC, depicted in Figure \ref{fig:HUNCC_OVERVIEW} and introduced in \cite{cohen2021network}. Here we provide a detailed explanation of the HUNCC algorithm \cite{cohen2021network}, as outlined in Algorithm \ref{alg:hybrid_scheme}, which describes HUNCC between a sender, Alice, and a receiver, Bob.

Starting with the encoding process at Alice's end, let $(\mathrm{Enc}, \mathrm{Dec}, p_k, s_k)$ represent a cryptosystem, as defined in Definition \ref{cryptosystem_defn}, with a security level $b$. Alice selects $c$ input channels for encryption. Without loss of generality, assume the channels indexed by $1, \ldots, c$ are the encrypted ones. The total number of channels, $n$, is fixed, and we consider message blocks $M^{(1)}, \ldots, M^{(N)}$, where each $M^{(i)} = [M_1^{(i)}, \ldots, M_n^{(i)}]$ with $M_j^{(i)} \in \mathbb{F}_{2^n}$ are independently and uniformly generated at random. Let $\mathbf{G} \in \mathbb{F}_{2^m}^{n \times n}$ be a Maximum Rank Distance (MRD) secure linear code, as detailed in \cite{cohen2018secure, cohen2021network}. Consequently, the vectors $X^{(1)}, \ldots, X^{(N)}$, where $X^{(i)} = \mathbf{G} M^{(i)}$, represent the MRD secrecy encoding of $M^{(i)}$ (refer to lines 1-5 in Algorithm \ref{alg:hybrid_scheme}).

For each encrypted input channel $i \in \{1, \ldots, c\}$, the symbols $X_i^{(1)}, \ldots, X_i^{(N)}$ are processed. Since $\mathbb{F}_{2^n} \simeq \mathbb{F}_2^n$, these symbols are injectively mapped into a bit sequence $\ddot{\mathbf{b}}_i$ of length $k_{in}$. Each $\ddot{\mathbf{b}}_i$ is encrypted using a key before transmission, resulting in $\mathbf{y}_i = \mathrm{Enc}(\ddot{\mathbf{b}}_i, p_i)$, with $\mathbf{y}_i$ having length $n$ (see lines 6-13 in Algorithm \ref{alg:hybrid_scheme}). For the $n-c$ unencrypted channels $i > c$, Alice directly transmits the symbols $X_i^{(1)}, \ldots, X_i^{(N)}$ (lines 14-16 in Algorithm \ref{alg:hybrid_scheme}).

The decoding process at Bob's receiving end is the following: For the $c$ encrypted channels, Bob uses the private keys to decrypt the received messages (lines 17-20 in Algorithm \ref{alg:hybrid_scheme}), reconstructing $[X_i^{(1)}, \ldots, X_i^{(N)}] = \mathrm{Dec}(\mathbf{y}_i, s_i)$ for $i \in \{1, \ldots, c\}$. The remaining $n-c$ channels, which were transmitted unencrypted, are directly available. Combining these with the decrypted messages from the $c$ encrypted channels, Bob obtains the complete set $X^{(1)}, \ldots, X^{(N)}$. Finally, for each column of the reconstructed encoded block, Bob applies the parity check matrix $\mathbf{H} \in \mathbb{F}_{2^m}^{n \times n}$ to recover the original transmitted messages (lines 21-24 in Algorithm \ref{alg:hybrid_scheme}).

HUNCC's approach blends information-theoretic and computational security to enable efficient and secure encryption capable of handling high data rates \cite[Theorem 3]{cohen2021network}. Notably, HUNCC maintains the computational security of the underlying cryptosystem even when only part of the data is encrypted \cite[Theorem 1]{cohen2021network}. This reduces computational demands and significantly lowers energy consumption while preserving strong security. Its flexibility across various communication networks and cryptosystems makes it ideal for resource-constrained environments. Algorithm \ref{alg:hybrid_scheme} outlines HUNCC in a communication context. The cryptanalysis performed on HUNCC in section \ref{results}, further validates the encryption mechanism by providing comparable empirical bounds to that of other widely used encryption schemes.

\subsection{Indistinguishability for Individual Security}\label{IndividualINDCPA_DEFN}

To analyze our network coding based schemes for individual security, we used a modified IND-CPA experiment, where we test for individual security by only choosing one of the outgoing channels/individual messages, as introduced in \cite{cohen2022partial}. Individual indistinguishability under Chosen Plaintext Attack (Individual IND-CPA) can be described as the following below:

\textbf{\textit{Individual IND-CPA}:} Let the set of messages be $\mathcal{M} = \mathbb{F}_{q}^{k_u}$, with each message going through an individual channel. Thus, each message $m = (m_1, \ldots, m_{k_u})$. We refer to each $m_i$ as an individual message. Then, Individual IND-CPA is defined by the following game between an adversary \(\mathcal{A}\) and a challenger \(\mathcal{C}\):
\begin{itemize}
    \item The challenger \(\mathcal{C}\) generates a key pair $\text{Gen}(\kappa) = (pk, sk)$ for some security parameter $\kappa$ and shares the public key $pk$ with the adversary \(\mathcal{A}\).
    \item \(\mathcal{A}\) may send a polynomial amount of plaintexts to the challenger and receive back their encryptions. They may also perform a polynomial amount of operations.
    \item The adversary \(\mathcal{A}\) chooses an index $j^* \in \{1, \ldots, k_u\}$ and two challenge individual messages $m_{j^*}^1$ and $m_{j^*}^2$, and sends them to the challenger.
    \item The challenger chooses $i \in \{0, 1\}$ uniformly at random.
    \item The challenger chooses $k_u - 1$ individual messages $m_j$, for $j \in \{1, \ldots, k_u\} - \{j^*\}$ uniformly at random and then constructs the message $m = (m_1, \ldots, m_{k_u})$, where $m_{j^*} = m_{j^*}^i$.
    \item The challenger \(\mathcal{C}\) sends the challenge ciphertext $c^* = \text{Enc}(m, pk)$ to the adversary.
    \item The adversary performs a polynomial amount of operations before outputting a guess $b'$ for $b$. The experiment outputs 1 if the guess is correct.
\end{itemize}

The cryptosystem is individually indistinguishable under chosen plaintext attack if any adversary has only a negligible advantage over a uniformly random guess of $b$, i.e.,
$$Pr [\text{Individual\_CPA}_{\text{Game}}^{\mathcal{A}} = 1] \leq 1/2 + f(\kappa)$$
where $f(\kappa)$ is a negligible function.

One can see that this is very similar to IND-CPA, defined in \ref{INDCPA_defn}. This modified version of IND-CPA provides a computational analogue to individual security \cite{cohen2022partial}. It guarantees that an adversary can learn only a negligible amount of information from any individual channel, but is suitable for a setting where an adversary can see all the channels.

\bibliographystyle{IEEEtran}
\bibliography{references.bib}

\begin{thebibliography}{10}
\providecommand{\url}[1]{#1}
\csname url@samestyle\endcsname
\providecommand{\newblock}{\relax}
\providecommand{\bibinfo}[2]{#2}
\providecommand{\BIBentrySTDinterwordspacing}{\spaceskip=0pt\relax}
\providecommand{\BIBentryALTinterwordstretchfactor}{4}
\providecommand{\BIBentryALTinterwordspacing}{\spaceskip=\fontdimen2\font plus
\BIBentryALTinterwordstretchfactor\fontdimen3\font minus \fontdimen4\font\relax}
\providecommand{\BIBforeignlanguage}[2]{{%
\expandafter\ifx\csname l@#1\endcsname\relax
\typeout{** WARNING: IEEEtran.bst: No hyphenation pattern has been}%
\typeout{** loaded for the language `#1'. Using the pattern for}%
\typeout{** the default language instead.}%
\else
\language=\csname l@#1\endcsname
\fi
#2}}
\providecommand{\BIBdecl}{\relax}
\BIBdecl

\bibitem{kim2024crypto}
B.~D. Kim, V.~A. Vasudevan, J.~Woo, A.~Cohen, R.~G. D’Oliveira, T.~Stahlbuhk, and M.~M{\'e}dard, ``Crypto-mine: Cryptanalysis via mutual information neural estimation,'' in \emph{ICASSP 2024-2024 IEEE International Conference on Acoustics, Speech and Signal Processing (ICASSP)}.\hskip 1em plus 0.5em minus 0.4em\relax IEEE, 2024, pp. 4820--4824.

\bibitem{mavroeidis2018impact}
V.~Mavroeidis, K.~Vishi, M.~D. Zych, and A.~J{\o}sang, ``The impact of quantum computing on present cryptography,'' \emph{arXiv preprint arXiv:1804.00200}, 2018.

\bibitem{chen2016report}
L.~Chen, L.~Chen, S.~Jordan, Y.-K. Liu, D.~Moody, R.~Peralta, R.~A. Perlner, and D.~Smith-Tone, \emph{Report on post-quantum cryptography}.\hskip 1em plus 0.5em minus 0.4em\relax US Department of Commerce, National Institute of Standards and Technology~…, 2016, vol.~12.

\bibitem{bos2018crystals}
J.~Bos, L.~Ducas, E.~Kiltz, T.~Lepoint, V.~Lyubashevsky, J.~M. Schanck, P.~Schwabe, G.~Seiler, and D.~Stehl{\'e}, ``Crystals-kyber: a cca-secure module-lattice-based kem,'' in \emph{2018 IEEE European Symposium on Security and Privacy (EuroS\&P)}.\hskip 1em plus 0.5em minus 0.4em\relax IEEE, 2018, pp. 353--367.

\bibitem{prest2020falcon}
T.~Prest, P.-A. Fouque, J.~Hoffstein, P.~Kirchner, V.~Lyubashevsky, T.~Pornin, T.~Ricosset, G.~Seiler, W.~Whyte, and Z.~Zhang, ``Falcon,'' \emph{Post-Quantum Cryptography Project of NIST}, 2020.

\bibitem{dubrova2023breaking}
E.~Dubrova, K.~Ngo, J.~G{\"a}rtner, and R.~Wang, ``Breaking a fifth-order masked implementation of crystals-kyber by copy-paste,'' in \emph{Proceedings of the 10th ACM Asia Public-Key Cryptography Workshop}, 2023, pp. 10--20.

\bibitem{wenger2022salsa}
E.~Wenger, M.~Chen, F.~Charton, and K.~E. Lauter, ``Salsa: Attacking lattice cryptography with transformers,'' \emph{Advances in Neural Information Processing Systems}, vol.~35, pp. 34\,981--34\,994, 2022.

\bibitem{stinson2018cryptography}
D.~R. Stinson and M.~Paterson, \emph{Cryptography: Theory and Practice}, 4th~ed.\hskip 1em plus 0.5em minus 0.4em\relax Chapman and Hall/CRC Press, 2018.

\bibitem{gohr2019improving}
A.~Gohr, ``Improving attacks on round-reduced speck32/64 using deep learning,'' in \emph{Advances in Cryptology--CRYPTO 2019: 39th Annual International Cryptology Conference, Santa Barbara, CA, USA, August 18--22, 2019, Proceedings, Part II 39}.\hskip 1em plus 0.5em minus 0.4em\relax Springer, 2019, pp. 150--179.

\bibitem{dwork2006differential}
C.~Dwork, ``Differential privacy,'' in \emph{International colloquium on automata, languages, and programming}.\hskip 1em plus 0.5em minus 0.4em\relax Springer, 2006, pp. 1--12.

\bibitem{dwork2008differential}
------, ``Differential privacy: A survey of results,'' in \emph{International conference on theory and applications of models of computation}.\hskip 1em plus 0.5em minus 0.4em\relax Springer, 2008, pp. 1--19.

\bibitem{tang2017privacy}
J.~Tang, A.~Korolova, X.~Bai, X.~Wang, and X.~Wang, ``Privacy loss in apple's implementation of differential privacy on macos 10.12,'' \emph{arXiv preprint arXiv:1709.02753}, 2017.

\bibitem{near2018differential}
J.~Near, ``Differential privacy at scale: Uber and berkeley collaboration,'' in \emph{Enigma 2018 (Enigma 2018)}, 2018.

\bibitem{abowd2018us}
J.~M. Abowd, ``The us census bureau adopts differential privacy,'' in \emph{Proceedings of the 24th ACM SIGKDD international conference on knowledge discovery \& data mining}, 2018, pp. 2867--2867.

\bibitem{nasr2023tight}
M.~Nasr, J.~Hayes, T.~Steinke, B.~Balle, F.~Tram{\`e}r, M.~Jagielski, N.~Carlini, and A.~Terzis, ``Tight auditing of differentially private machine learning,'' in \emph{32nd USENIX Security Symposium (USENIX Security 23)}, 2023, pp. 1631--1648.

\bibitem{steinke2024privacy}
T.~Steinke, M.~Nasr, and M.~Jagielski, ``Privacy auditing with one (1) training run,'' \emph{Advances in Neural Information Processing Systems}, vol.~36, 2024.

\bibitem{shimeall2013introduction}
T.~Shimeall and J.~Spring, \emph{Introduction to information security: a strategic-based approach}.\hskip 1em plus 0.5em minus 0.4em\relax Newnes, 2013.

\bibitem{belghazi2018mine}
M.~I. Belghazi, A.~Baratin, S.~Rajeswar, S.~Ozair, Y.~Bengio, A.~Courville, and R.~D. Hjelm, ``Mine: mutual information neural estimation,'' \emph{arXiv preprint arXiv:1801.04062}, 2018.

\bibitem{zhang2022privacy}
W.~Zhang, B.~Jiang, M.~Li, and X.~Lin, ``Privacy-preserving aggregate mobility data release: An information-theoretic deep reinforcement learning approach,'' \emph{IEEE Transactions on Information Forensics and Security}, vol.~17, pp. 849--864, 2022.

\bibitem{razeghi2023bottlenecks}
B.~Razeghi, F.~P. Calmon, D.~Gunduz, and S.~Voloshynovskiy, ``Bottlenecks club: Unifying information-theoretic trade-offs among complexity, leakage, and utility,'' \emph{IEEE Transactions on Information Forensics and Security}, vol.~18, pp. 2060--2075, 2023.

\bibitem{rodriguez2021variational}
B.~Rodr{\'\i}guez-G{\'a}lvez, R.~Thobaben, and M.~Skoglund, ``A variational approach to privacy and fairness,'' in \emph{2021 IEEE Information Theory Workshop (ITW)}.\hskip 1em plus 0.5em minus 0.4em\relax IEEE, 2021, pp. 1--6.

\bibitem{esfahanizadeh2023infoshape}
H.~Esfahanizadeh, W.~Wu, M.~Ghobadi, R.~Barzilay, and M.~M{\'e}dard, ``Infoshape: Task-based neural data shaping via mutual information,'' in \emph{ICASSP 2023-2023 IEEE International Conference on Acoustics, Speech and Signal Processing (ICASSP)}.\hskip 1em plus 0.5em minus 0.4em\relax IEEE, 2023, pp. 1--5.

\bibitem{kale2024texshape}
H.~K. Kale, H.~Esfahanizadeh, N.~Elias, O.~Baser, M.~Medard, and S.~Vishwanath, ``Texshape: Information theoretic sentence embedding for language models,'' \emph{arXiv preprint arXiv:2402.05132}, 2024.

\bibitem{bellare2003theoretical}
M.~Bellare and T.~Kohno, ``A theoretical treatment of related-key attacks: Rka-prps, rka-prfs, and applications,'' in \emph{International Conference on the Theory and Applications of Cryptographic Techniques}.\hskip 1em plus 0.5em minus 0.4em\relax Springer, 2003, pp. 491--506.

\bibitem{standaert2010introduction}
F.-X. Standaert, ``Introduction to side-channel attacks,'' \emph{Secure integrated circuits and systems}, pp. 27--42, 2010.

\bibitem{randolph2020power}
M.~Randolph and W.~Diehl, ``Power side-channel attack analysis: A review of 20 years of study for the layman,'' \emph{Cryptography}, vol.~4, no.~2, p.~15, 2020.

\bibitem{liang2009information}
Y.~Liang, H.~V. Poor, and S.~Shamai, \emph{Information theoretic security}.\hskip 1em plus 0.5em minus 0.4em\relax Now Publishers Inc, 2009.

\bibitem{shannon1949communication}
C.~E. Shannon, ``Communication theory of secrecy systems,'' \emph{The Bell system technical journal}, vol.~28, no.~4, pp. 656--715, 1949.

\bibitem{goodfellow2016deep}
I.~Goodfellow, Y.~Bengio, and A.~Courville, \emph{Deep learning}.\hskip 1em plus 0.5em minus 0.4em\relax MIT press, 2016.

\bibitem{choi2020regularized}
K.~Choi and S.~Lee, ``Regularized mutual information neural estimation,'' 2020.

\bibitem{cachin1997entropy}
C.~Cachin, ``Entropy measures and unconditional security in cryptography,'' Ph.D. dissertation, ETH Zurich, 1997.

\bibitem{standard1999data}
D.~E. Standard \emph{et~al.}, ``Data encryption standard,'' \emph{Federal Information Processing Standards Publication}, vol. 112, p.~3, 1999.

\bibitem{daemen1999aes}
J.~Daemen and V.~Rijmen, ``Aes proposal: Rijndael,'' 1999.

\bibitem{rivest1978method}
R.~L. Rivest, A.~Shamir, and L.~Adleman, ``A method for obtaining digital signatures and public-key cryptosystems,'' \emph{Communications of the ACM}, vol.~21, no.~2, pp. 120--126, 1978.

\bibitem{cohen2021network}
A.~Cohen, R.~G.~L. D’Oliveira, S.~Salamatian, and M.~M{\'e}dard, ``Network coding-based post-quantum cryptography,'' \emph{IEEE journal on selected areas in information theory}, vol.~2, no.~1, pp. 49--64, 2021.

\bibitem{shannon1945mathematical}
C.~E. Shannon, ``A mathematical theory of cryptography,'' \emph{Bell System Technical Memo MM 45-110-02}, 1945.

\bibitem{bellare1995optimal}
M.~Bellare and P.~Rogaway, ``Optimal asymmetric encryption,'' in \emph{Advances in Cryptology—EUROCRYPT'94: Workshop on the Theory and Application of Cryptographic Techniques Perugia, Italy, May 9--12, 1994 Proceedings 13}.\hskip 1em plus 0.5em minus 0.4em\relax Springer, 1995, pp. 92--111.

\bibitem{cohen2022partial}
A.~Cohen, R.~G.~L. D’Oliveira, K.~R. Duffy, and M.~M{\'e}dard, ``Partial encryption after encoding for security and reliability in data systems,'' in \emph{2022 IEEE International Symposium on Information Theory (ISIT)}.\hskip 1em plus 0.5em minus 0.4em\relax IEEE, 2022, pp. 1779--1784.

\bibitem{woo2023cermet}
J.~Woo, V.~A. Vasudevan, B.~Kim, A.~Cohen, R.~G.~L. D'Oliveira, T.~Stahlbuhk, and M.~M{\'e}dard, ``{CERMET: Coding for Energy Reduction with Multiple Encryption Techniques-- It's easy being green},'' \emph{arXiv preprint arXiv:2308.05063}, 2023.

\bibitem{10619109}
S.~Tarnopolsky and A.~Cohen, ``Coding-based hybrid post-quantum cryptosystem for non-uniform information,'' in \emph{2024 IEEE International Symposium on Information Theory (ISIT)}, 2024, pp. 1830--1835.

\bibitem{hjelm2018learning}
R.~D. Hjelm, A.~Fedorov, S.~Lavoie-Marchildon, K.~Grewal, P.~Bachman, A.~Trischler, and Y.~Bengio, ``Learning deep representations by mutual information estimation and maximization,'' \emph{arXiv preprint arXiv:1808.06670}, 2018.

\bibitem{kambhatla2018decipherment}
N.~Kambhatla, ``Decipherment of substitution ciphers with neural language models,'' 2018.

\bibitem{NIST-PQC}
\BIBentryALTinterwordspacing
NIST, ``Post quantum cryptography,'' NIST Computer Security Resource Center, Tech. Rep., 2017. [Online]. Available: \url{https://csrc.nist.gov/projects/post-quantum-cryptography/post-quantum-cryptography-standardization/evaluation-criteria/security-(evaluation-criteria)}
\BIBentrySTDinterwordspacing

\bibitem{cohen2018secure}
A.~Cohen, A.~Cohen, M.~Medard, and O.~Gurewitz, ``Secure multi-source multicast,'' \emph{IEEE Transactions on Communications}, vol.~67, no.~1, pp. 708--723, 2018.

\end{thebibliography}

\end{document}